\documentclass[aps,prd,twocolumn,nofootinbib,showpacs,showkeys,superscriptaddress]{revtex4-1}

\usepackage{psfrag}
\usepackage{mathrsfs}
\usepackage{amssymb, bm}
\usepackage{amsmath, amsthm}
\usepackage{epstopdf}
\usepackage{hyperref}
\usepackage{enumerate}
\usepackage{longtable}
\usepackage{subfigure}
\usepackage{color}
\usepackage{mathrsfs}
\usepackage{graphicx}
\usepackage{bm,natbib,url,textcase}
\usepackage[english]{babel}
\usepackage[T1]{fontenc}
\usepackage{comment}
\numberwithin{equation}{section}

\usepackage{geometry}
\allowdisplaybreaks

\hypersetup{
    colorlinks=true,        
    linkcolor=blue,         
    citecolor=blue,         
    urlcolor=blue           
}
\usepackage{tikz}

\newcommand{\be}{\begin{equation}}
\newcommand{\ee}{\end{equation}}

\definecolor{purple}{rgb}{1,0,1}
\definecolor{lime}{HTML}{a6CE39} 

\newcommand{\orcidicon}{%
	\begin{tikzpicture}
		\draw[lime, fill=lime] (0,0) 
		circle [radius=0.15] 
		node[white] {{\fontfamily{qag}\selectfont \tiny ID}};
		\draw[white, fill=white] (-0.0625,0.095) 
		circle [radius=0.007];
	\end{tikzpicture}	\hspace{-2mm}
}
\newcommand\orcidMarcello{{\href{https://orcid.org/0000-0003-0397-2705}{\orcidicon}}}
\newcommand\orcidDaniele{{\href{https://orcid.org/0000-0003-4379-2549}{\orcidicon}}}
\newcommand\orcidSalvatore{{\href{https://orcid.org/0000-0003-4886-2024}{\orcidicon}}}
\newcommand\orcidValerio{{\href{https://orcid.org/0000-0002-2601-1870}{\orcidicon}}}

\def\av{``}                    
\def\cv{'' }                   
\def\z{~}                      
\def\nn{\nonumber}             

\begin{document}
\def\theequation{\arabic{section}.\arabic{equation}}

\title{Generalized McVittie geometry in Horndeski gravity with matter}

\author{Marcello Miranda\orcidMarcello}
\email{marcello.miranda@unina.it}
\affiliation{Scuola Superiore Meridionale, Largo San Marcellino 10, 
I-80138, Napoli, Italy}
\affiliation{Dipartimento di Fisica ``E. Pancini'', Universit\`{a} di 
Napoli ``Federico II'', Napoli, Italy}
\affiliation{INFN Sez. di Napoli, Compl. Univ. Monte S. Angelo, 
Edificio G, Via	Cinthia, I-80126, Napoli, Italy}

\author{Daniele Vernieri\orcidDaniele}
\email{daniele.vernieri@unina.it}
\affiliation{Dipartimento di Fisica ``E. Pancini'', Universit\`{a} di 
Napoli ``Federico II'', Napoli, Italy}
\affiliation{INFN Sez. di Napoli, Compl. Univ. Monte S. Angelo, 
Edificio G, Via	Cinthia, I-80126, Napoli, Italy}

\author{Salvatore Capozziello\orcidSalvatore}
\email{capozziello@na.infn.it}
\affiliation{Scuola Superiore Meridionale, Largo San Marcellino 10, 
I-80138, Napoli, Italy}
\affiliation{Dipartimento di Fisica ``E. Pancini'', Universit\`{a} di 
Napoli ``Federico II'', Napoli, Italy}
\affiliation{INFN Sez. di Napoli, Compl. Univ. Monte S. Angelo, 
Edificio G, Via	Cinthia, I-80126, Napoli, Italy}

\author{Valerio Faraoni\orcidValerio}
\email{vfaraoni@ubishops.ca}
\affiliation{Department of Physics \& Astronomy, Bishop's University, 2600 College Street, Sherbrooke, Qu\'ebec, Canada J1M~1Z7}

\begin{abstract}

We investigate McVittie and generalized McVittie solutions for Horndeski gravity with a spatially homogeneous gravitational scalar field, which is stealth at small scales near the central object but, at large scales, sources the FLRW universe in which the central inhomogeneity is embedded. Unlike previous studies, we include matter and obtain generalized McVittie solutions in the extended cuscuton model. The possible configurations are classified according to the time-dependence of the gravitational coupling, the radial energy flow, the accretion rate onto the central object, and the Hubble rate.

\end{abstract}

\date{\today}

\maketitle

\section{Introduction}\label{sec:1}
\setcounter{equation}{0}

The McVittie solution of the Einstein field equations, sourced by a fluid\z\cite{McVittie, Stephani, Krasinski, AHbook, 
Faraoni:2021nhi}, describes a 
central object embedded in a Friedmann--Lema\^itre--Robertson--Walker (FLRW) 
universe and it is a special case of the Kustaanheimo--Qvist family of 
shear-free solutions\z\cite{KustaanheimoQvist}. McVittie chose to forbid 
explicitly the accretion of cosmic fluid onto the central object\z\cite{McVittie}. This cosmic fluid has a homogeneous energy density 
$\rho(t)$ and an inhomogeneous, radially-dependent, pressure $P(t,r)$ 
(unless the cosmic fluid is replaced by a cosmological constant, in which 
case the McVittie solution degenerates into a Schwarzschild--de 
Sitter/Kottler geometry\z\cite{McVittie}). In addition to relevant 
previous literature, in the past decade, the McVittie 
solution  has been the subject of much work,  
mostly devoted to understand the nature of the time-dependent 
apparent horizon  covering a singularity at a finite radius\z\cite{NewmanMcVittie82, Sussman85, NIS, Ferrarisetal, Nolan:1998xs, 
Nolan99CQG, GaoZhang04, Gao:2004gz, Carrera:2009ve, Kaloper:2010ec, 
Lake:2011ni, Anderson11, Nandra:2011ug, Nandra:2011ui, Arakida:2011st, 
Guariento:2012ri, Faraoni:2012gz, Landry:2012nv, LeDelliou:2013qmm, 
Faraoni:2012je, daSilvaFontaniniGuariento13}. This apparent horizon has 
been shown to depend on the form of the scale factor $a(t)$ of the FLRW 
universe where the object is embedded\z\cite{daSilvaFontaniniGuariento13}. The McVittie geometry has also been used  
to study the interplay between local dynamics and cosmological 
expansion\z\cite{Faraoni:2007es, Carrera:2009ve}, to model cosmological 
wormholes\z\cite{Faraoni:2005it}, and generalized McVittie geometries  
allowing for cosmic fluid accretion have been introduced\z\cite{Faraoni:2007es} and studied\z\cite{Gao:2008jv, Faraoni:2008tx, Carrera:2009ve}.

Since a minimally coupled scalar field is equivalent to an irrotational 
perfect fluid, it is natural to ask whether a scalar field can  
source the McVittie metric in Einstein gravity. The answer is negative but 
a gravitational scalar field, 
nonminimally coupled to the curvature (a signature of alternative gravity) has been shown to be a possible 
effective source for  McVittie and generalized McVittie geometries\z\cite{Afshordi:2006ad,Abdalla:2013ara}. 
This context is different: theories of gravity alternative to general 
relativity have seen a huge resurgence of interest with the goal of  
explaining the current 
acceleration of the universe discovered in 1998 with type Ia supernovae 
without advocating an {\em ad hoc} dark energy\z\cite{book,Bamba,Rept}. It is then natural to ask 
whether scalar-tensor gravity, possibly in its most general formulations 
given by Horndeski and DHOST gravity, admits (generalized)  
McVittie solutions. This question was answered affirmatively in 
Refs.\z\cite{Afshordi:2006ad,Abdalla:2013ara, 
Afshordi:2006ad,Abdalla:2013ara}, but there are severe restrictions on 
the class of theories in which this is possible. McVittie spacetimes are also 
exact solutions 
of shape dynamics  of interest for quantum gravity\z\cite{shapedynamics}, 
and are non-deformable solutions for extended  $f(T)$ teleparallel gravity, where $T$ is the 
torsion scalar (\cite{BejaranoFerraroGuzman17}, see also Ref.\z\cite{Cai} for a general discussion on teleparallel gravity).

It is  time to revisit McVittie and generalized McVittie spacetimes in the most general 
scalar-tensor theories for three reasons. First, our understanding of the cuscuton that admits McVittie geometries as solutions has improved in the last decade and, very recently, the ``extended cuscuton'' has been introduced in research devoted to avoid cosmological singularities\z\cite{Iyonaga:2018vnu, Iyonaga:2020bmm, Quintin:2019orx}. Second, the previous analyses were restricted to vacuum but, ultimately, a realistic description of a cosmological setting must include matter, which is what we do in this article. Third, there has been growing attention to stealth solutions of scalar-tensor gravity with motivations coming mainly from the possibility that black hole hair arising in stealth solutions  could be detected directly in gravitational wave observations. Such a a detection would discriminate between general relativity and scalar-tensor gravity. Both ``first generation'' scalar-tensor and more 
modern Horndeski theories admit stealth solutions  
that circumvent the no-hair theorems and for which the scalar field does 
not gravitate. They include the Schwarzschild geometry with a 
non-vanishing scalar field $\phi = \phi_0 t$\,. Then, it makes sense to look 
for ``locally stealth'' McVittie solutions of Horndeski theories with a 
gravitational scalar field, the geometry being McVittie and the scalar 
being homogeneous, $ \phi = \phi (t)$\,. This scalar field would be a 
stealth with respect to the central object, but would contribute to 
creating the FLRW ``background'' (indeed, in the first part of 
the present work, it will be the only effective cosmic fluid). Horndeski 
gravity is the most general scalar-tensor theory with second order 
equations of motion, which automatically avoids the notorious Ostrogradsky 
instability encountered for higher order field equations\z\cite{Ostrogradsky,Woodard:2015zca}. Continuing, the generalized McVittie geometries with a radial energy flow offer more possibilities for modelling, but have not yet received sufficient attention because of their higher complexity with respect to ``standard'' McVittie geometries. 

Last but not least, McVittie and generalized McVittie spacetimes offer toy models for cosmological inhomogeneities which can be exploited to investigate the current problem of  Hubble tension in relation with local inhomogeneities\z\cite{Riess:2019qba, Castelvecchi, DiValentino:2021izs, Spallicci}.

To fix the ideas, we follow the notation of Ref.\z\cite{Waldbook}, using 
units in which the speed of light $c$ and Newton's constant $G$ are unity. 
The metric signature is ${-}{+}{+}{+}$, $ R_{\mu\nu}$ denotes the 
Ricci tensor, $ R\equiv g^{\mu\nu} R_{\mu\nu}$ is the Ricci 
scalar, $G_{\mu\nu} \equiv  R_{\mu\nu}-\frac{1}{2}\, g_{\mu\nu} 
 R$ 
is the Einstein tensor, while $g$ is the determinant of the metric tensor 
$g_{\mu\nu}$ and round brackets around indices denote symmetrization.

The action of Horndeski gravity\z\cite{Horndeski} is written as\z\cite{Kobayashi:2019hrl, Capozziello:2018gms}
\be
S\left[ g_{\mu\nu}, \phi \right] = \int d^4 x \sqrt{-g} \, \left( {\cal 
L}_2 + 
{\cal L}_3+ {\cal L}_4+ 
{\cal L}_5 \right) + S^\mathrm{(matter)} \,, \label{Horndeskiaction}
\ee
where 
\begin{eqnarray}
{\cal L}_2 &=& G_2\left( \phi, X \right) \,,\\  
&&\nonumber\\
{\cal L}_3 &=& - G_3\left( \phi, X \right) \Box \phi \,,\\  
&&\nonumber\\
{\cal L}_4 &=& G_4\left( \phi, X \right)  R +G_{4,X} \left( \phi, X 
\right) 
\left[ \left( \Box \phi \right)^2 -\left( \nabla_{\mu} \nabla_{\nu} \phi 
\right)^2 \right]  \,, \nonumber\\  
&&\\
{\cal L}_5 & = & G_5\left( \phi, X \right) G_{\mu\nu}\nabla^{\mu} 
\nabla^{\nu} 
\phi \nonumber\\
&&\nonumber\\
&\, & -\frac{ G_{5,X} }{6} \left[ \left( \Box\phi \right)^3 -3\Box\phi 
\left( \nabla_{\mu} \nabla_{\nu} \phi \right)^2 +2\left( \nabla_{\mu} 
\nabla_{\nu} \phi \right)^3 \right] \,,\nonumber\\
&&
\end{eqnarray}
where $\phi$ is the gravitational scalar field degree of freedom, $ X 
\equiv 
-\nabla^{\alpha}\phi  \nabla_{\alpha} \phi /2 $, the $G_i\left(\phi, 
X\right)$  ($i=2,3,4,5$) are regular functions of 
$\phi$ and $X$, while $ G_{i,\phi} \equiv \partial G_i/\partial \phi$ and 
$ 
G_{i,X} \equiv \partial G_i/\partial X$\,. Moreover, 
$ \left( \nabla_{\mu} \nabla_{\nu} \phi \right)^2 \equiv 
\nabla_{\mu} \nabla_{\nu} \phi 
\nabla^{\mu} \nabla^{\nu} \phi$ and 
$ \left( \nabla_{\mu} \nabla_{\nu} \phi \right)^3 \equiv 
\nabla_{\mu} \nabla_{\sigma} \phi \nabla^{\sigma} \nabla^{\alpha} \phi  
\nabla_{\alpha} \nabla^{\mu} \phi $\,.

The multi-messenger event GW170817/GRB170817~A from a binary neutron star merger\z\cite{LIGOScientific:2017vwq, LIGOScientific:2017zic} 
essentially ruled out 
theories in which the speed of gravitational waves differs from $c$\,. The 
surviving Horndeski theories are also those that avoid  
instabilities and admit an Einstein frame description and satisfy
\z\cite{Bettoni:2016mij, Creminelli:2018xsv, Kobayashi:2019hrl}
\be
G_5=0 \,, \quad\quad  G_{4,X}=0 \,. \label{viabilityconds}
\ee
We restrict ourselves to this subclass, for which the  
Lagrangian density reduces to 
\begin{equation} \label{action}
{\cal L} =  
G_2(\phi,X) - G_3(\phi,X) \Box \phi + G_4(\phi)  R  \,. 
\end{equation}
The corresponding field equations are
\begin{widetext}
\begin{eqnarray}
&&  G_4(\phi) \, G_{\mu\nu} -\nabla_{\mu}\nabla_{\nu}G_4(\phi) 
+ \left[ \Box G_4(\phi)-\dfrac{G_2 (\phi,X)}{2} 
-\dfrac{1}{2} \, \nabla_{\lambda}  
\phi\nabla^{\lambda}G_{3}(\phi,X)
\right] g_{\mu\nu}     \nonumber\\
&&\nonumber\\
&& + \left[ \dfrac{G_{3,X}(\phi,X)}{2} \, 
\Box\phi   -\dfrac{G_{2,X}(\phi,X)}{2} \right]  
\nabla_{\mu}\phi\nabla_{\nu}\phi  
+ \nabla_{(\mu}\phi \nabla_{\nu)}G_{3}(\phi,X)=0   \label{hfeq}
\end{eqnarray}
and
\begin{equation}\label{heom}
\begin{array}{ll}
& G_{4,\phi}(\phi)  R + G_{2,\phi}(\phi,X)+G_{2,X}(\phi,X) 
\Box\phi+\nabla_{\lambda}\phi\nabla^{\lambda}G_{2,X}(\phi,X)\vspace{7pt}\\
&-G_{3,X}(\phi,X)(\Box\phi)^2-\nabla_{\lambda}\phi\nabla^\lambda 
G_{3,X} (\phi,X)\Box\phi-G_{3,X}(\phi,X)\nabla^{\lambda} 
\phi\Box\nabla_\lambda\phi+G_{3,X}(\phi,X)  R_{\mu\nu}\nabla^{\mu} 
\phi\nabla^{\nu}\phi\vspace{7pt}\\
   &-\Box G_{3}(\phi,X)-G_{3,\phi}(\phi,X)\Box\phi=0 \,.
\end{array}
\end{equation}
\end{widetext}
We search for McVittie solutions of these field 
equations.

The layout of the paper is the following. Sec.\z\ref{sec:2} is devoted to a general discussion of McVittie geometry in Horndeski theory. In particular, we discuss the McVittie scalar field as a cuscuton. Generalized McVittie solutions in Horndeski gravity are discussed in Sec.\z\ref{sec:3}, with attention to various cases with respect to the functions $G_2$, $G_3$, and $G_4$ entering the Lagrangian\z\eqref{action}. Horndeski-McVittie as an extended cuscuton model is studied in Sec.\z\ref{sec:4}, while McVittie solutions in extended cuscuton models in presence of matter is considered in Sec.\z\ref{sec:5} with various combinations of the above functions $G_i$\,. Conclusions are drawn in Sec.\z\ref{sec:10}.

\section{McVittie geometry in Horndeski theory}
\label{sec:2}
\setcounter{equation}{0}

Let us take into account the most general class of Horndeski theories in which the 
McVittie geometry is a solution of the field equations. The  McVittie line 
element in isotropic coordinates $\left( t, r, \vartheta, \varphi \right)$ 
is\z\cite{McVittie} 
\begin{eqnarray}\label{McVittie}
ds^2 &=& -\dfrac{\left(  1-\dfrac{m }{2 a(t) r} \right)^2}{\left( 1+\dfrac{m}{2 a(t) r}  \right)^2}\, dt^2 \nonumber\\
&&\nonumber\\
&\, &  +  a(t)^2\left(1+\dfrac{m}{2  
a(t) r }\right)^4\left(dr^2+r^2 d\Omega_{(2)}^2\right)\,,\label{mc}
\end{eqnarray} 
where the mass parameter $m$ is a positive constant, $a(t)$ is the scale factor of the FLRW 
``background'', and $ d\Omega_{(2)}^2 \equiv d\vartheta^2 +\sin^2 
\vartheta \, d\varphi^2$ is the line element on the unit 2-sphere. 
Following the recent literature, and in agreement with cosmological 
observations, we restrict ourselves to spatially flat 
FLRW universes. In the spirit of looking for ``locally stealth'' 
solutions discussed 
above, we consider a {\it homogeneous} scalar field $ \phi(t)$, for which 
\begin{equation}
\label{kin}
X (t, r) = \dfrac{1}{2} \, \left(\dfrac{ 2a(t)r+ m   }{2a(t)r -  m }\right)^2  \dot{\phi}^2(t)
\end{equation}
in the geometry\z(\ref{mc}), where an overdot denotes 
differentiation with respect to the comoving time $t$ of the FLRW 
``background''. The only independent field equations 
that are not identically satisfied are the $(t,t)$, $(t,r) $, $(r,r)$ and 
$(r,\theta)$ ones. The homogeneous scalar field $\phi(t)$ is equivalent to 
a {\em perfect} fluid. This is not true for a general 
Brans--Dicke-like or Horndeski scalar field 
which depends on the spatial position, which is instead equivalent to an 
{\em imperfect} fluid with heat conduction, shear and bulk viscosity, and
anisotropic stresses\z\cite{Pimentel89, Faraoni:2018qdr, Faraoni:2021lfc, 
Faraoni:2021jri, Giusti:2021sku}. The imperfect fluid structure is a 
consequence of the explicit nonminimal coupling of $\phi$ to gravity and 
disappears only in 
very special spacetimes, such as the FLRW ones where shear and currents  
would violate spatial isotropy\z\cite{Faraoni:2018qdr}, or in special theories (such as the cuscuton\z\cite{Quintin:2019orx}). The 
assumption  that the scalar $\phi$  is spatially homogeneous has the 
important consequence of making it 
possible for the McVittie solution to satisfy the Horndeski field 
equations. In fact, the latter are equivalent to effective Einstein 
equations with an extra effective fluid in their right-hand side and this 
fluid is dissipative if $\phi$ depends on the spatial position\z\cite{Pimentel89, Faraoni:2018qdr, Faraoni:2021lfc, Faraoni:2021jri, 
Giusti:2021sku}, while the 
McVittie geometry is the only perfect fluid solution of the Einstein 
equations which is simultaneously\footnote{The proof of this statement is the actual derivation of the line element~(\ref{mc}) in the original McVittie article\z\cite{McVittie}.} spherically symmetric, shear-free, and 
asymptotically FLRW with vanishing (radial) energy current\z\cite{Raychaudhuri} (see also Ref.\z\cite{Faraoni:2021nhi}). Therefore, if $\phi=\phi \left(t, \vec{x}\right)$, the 
corresponding effective fluid in scalar-tensor or Horndeski gravity is 
necessarily a dissipative one\z\cite{Pimentel89, Faraoni:2018qdr, 
Faraoni:2021lfc, Faraoni:2021jri, Giusti:2021sku} and the 
McVittie geometry cannot be a solution. 

In order to find the functional forms of $G_2,\,G_3$, and $G_4$ such that the McVittie geometry~(\ref{mc}) is a solution of Eqs.~(\ref{hfeq}) 
and~(\ref{heom}), we substitute the line element~(\ref{mc}) in  
Eq.~(\ref{hfeq}) and impose that $G_2,\,G_3$, and $G_4$ solve it.
The $(t,r)$ equation becomes 
\begin{equation}
\frac{4 m \, a(t) \, \dot{\phi}(t) }{ m^2-4a^2(t) r^2 } \left[ X(t, r) 
G_{3,X} (\phi,X) - G_{4,\phi}(\phi)\right] = 0\,, 
\end{equation}
and, assuming $ \dot{\phi}(t)\neq 0$, it gives
\be
G_{3,X}(\phi,X)= \frac{G_{4,\phi}(\phi)}{X}\,,
\end{equation}
which integrates to 
\begin{equation}\label{g3.0}
G_{3}(\phi,X)=G_{4,\phi}(\phi)\,{\rm ln} \, X +F(\phi)\,.
\end{equation}
Inserting this form of $G_{3}(\phi,X)$, the radial component 
of the Horndeski field equation~(\ref{hfeq}) yields 
\begin{eqnarray}
G_2(\phi,X) &=& 2 X \left[ G_{4,\phi\phi}(\phi) \left( {\rm 
ln} X -2\right) + F_{,\phi}(\phi)\right] \nonumber\\
&&\nonumber\\
&\, & - 4 \, \frac{2 a(t) r+m}{2 a(t) r-m} \, \dot{\phi}(t) \, H(t)  \,
G_{4,\phi}(\phi) \nonumber\\
&&\nonumber\\
&\, & + G_{4}(\phi) \left[ -6 H^2(t) - 4 \,  \dot{H} (t) \, \dfrac{2 r 
a(t)+m}{2 a(t) r-m} \right]  \,.\nonumber\\
&&  \label{g2.1}
\end{eqnarray}

Let us further assume that $\phi(t) $ is monotonic,  $ 
\dot{\phi}(t) \gtrless  0$, while $ m<2a(t)r$\,. Equation~(\ref{kin}) 
then yields
\begin{equation}
\dot{\phi}(t) = \pm \, \dfrac{2 a(t) r-m}{2 a(t) r+m} \,, \sqrt{2X(t, r)} 
\end{equation}
while Eq.~(\ref{g2.1}) becomes
\begin{eqnarray}
G_2^{\scriptscriptstyle{(\pm)}}(\phi,X) &=& 2 X  
\left[ G_{4,\phi\phi}(\phi) \Big( {\rm ln}  
X -2\Big)+F_{,\phi}(\phi)\right] \nonumber\\
&&\nonumber\\
&\, & \mp \, 4 \sqrt{2} \, \sqrt{X } \, H(t) \, G_{4,\phi}(\phi)\nonumber\\
&&\nonumber\\
&\, & + G_{4}(\phi)\left[ -6 H^2(t) \mp 4 \sqrt{2}\, \sqrt{X } \,\, \dfrac{ \dot{H}(t)}{ \dot{\phi}(t)}\right]\,.\nonumber\\
&&  \label{g2.2}
\end{eqnarray}
%
%
Since $\phi$ is monotonic, one can regard $H(t) \equiv  S(\phi)$ 
as a function of $\phi$ and then $
\dot{H}(t)=S_{,\phi}(\phi) \, \dot{\phi}(t)$\,. In order for 
$G_2^{\scriptscriptstyle{(\pm)}}(\phi,X)$ to depend only 
on $\phi(t)$ and $X(t, r)$, it must be 
\begin{eqnarray}
 G_2^{\scriptscriptstyle{(\pm)}}(\phi,X) &=& 2 X 
\left[ G_{4,\phi\phi}(\phi) \left( {\rm  
ln} X -2 \right) \right. \nonumber\\
&&\nonumber\\
&\, & \left.  +F_{,\phi}(\phi)\right] \mp 4 \sqrt{2} \, \sqrt{X} \,
S(\phi) \, G_{4,\phi}(\phi)  \nonumber\\
&&\nonumber\\
&\, & +  G_{4}(\phi) \left[-6 
S^2(\phi) \mp 4 \sqrt{2} \, \sqrt{X} \, S_{,\phi}(\phi) 
\right]\,.\nonumber\\
&&  \label{g2.3} 
\end{eqnarray}

Assuming the four-velocity 
\begin{equation} 
u^\mu_{(\phi)} \equiv \frac{\nabla^\mu\phi}{\sqrt{2X}} 
\ee
of the perfect fluid equivalent to the scalar field to be future-oriented, 
{\em i.e.},
 \begin{equation} 
u^0_{(\phi)} = - \left( \frac{ 2 a(t) r+m }{ 2 a(t) r-m } \right)^2 
\frac{  \dot{\phi}(t)}{\sqrt{2X}}>0 \,,
\ee
or
\be 
\dot{\phi}(t) = -\dfrac{2 \, a(t) r-m}{2 \, a(t) r+m} \, \sqrt{2X(t, r)} 
\,,
\end{equation}
the Horndeski functions that solve Eqs.\z(\ref{hfeq}) 
and\z(\ref{heom}) are 
\begin{eqnarray}
G_2(\phi,X) &=& 2 X \left[ G_{4,\phi\phi}(\phi) \left({\rm 
ln}  X -2\right)+F_{,\phi}(\phi)\right] \nonumber\\
&&\nonumber\\
&\, & +4 \sqrt{2X} \, S(\phi) \, G_{4,\phi}(\phi) \nonumber\\
&&\nonumber\\
&\, & + G_{4}(\phi) \left[ -6 S^2(\phi)  
+4\sqrt{2} \, \sqrt{X} \, S_{,\phi} (\phi)\right] \,,\nonumber\\ 
&& \label{g2}\\
&&\nn\\
G_{3}(\phi,X) &=& G_{4,\phi}(\phi)\,{\rm 
ln} X + F(\phi)\,, \label{g3}
\end{eqnarray}
where $ S(\phi)=H(t)$ and $G_4(\phi)$ are free functions. In general, 
$G_4(\phi)$ is completely unconstrained. Therefore, any 
set $\Big\{ G_4(\phi),\phi(t),m(t),a(t)\Big\}$ is a solution of the 
theory~(\ref{action}) for which Eqs.~(\ref{g2}) and\z(\ref{g3}) hold with 
$S(\phi)=H(t)$\,.

The function $F(\phi)$ appearing in Eqs.\z(\ref{g2}) and\z(\ref{g3}) can be safely neglected  because, when inserted in the action, it only appears in the total divergence $2 X\,F_{,\phi}(\phi)-F(\phi)\Box\phi = -\nabla_{\mu}[F(\phi)\nabla^{\mu}\phi]$, which generates a boundary term and does not contribute to the field equations.

If $m=0$, we obtain the same set of functions and $X$ reduces to  $X= 
\dot{\phi}^2(t) /2$\,. 

In the special situation $G_3 \equiv 0$ it is
$G_{4,\phi}(\phi)=0$ and $ 
G_4(\phi)=$~const.; then, as shown in the next section, the scalar field 
$\phi$ is not dynamical.

\subsection{McVittie scalar field as a cuscuton}
\label{k-ess}

Let us consider now the situation in which $G_4=1/2 $, which 
implies $G_3=0$ (keeping in mind that $F(\phi)$ only contributes a boundary term $-\int d^4 x \sqrt{-g}\, \nabla_{\mu} [F(\phi) \, \nabla^{\mu}\phi] $ to the action), according to Eq.~(\ref{g3}). 
Using Eqs.~(\ref{g3})  and~(\ref{g2}), the Horndeski action becomes $ S = 
S_g + S_\phi $, where 
\begin{eqnarray}
S_g & = & \frac{1}{2}\int d^4x \, \sqrt{-g} \,  R \,,\\
&&\nonumber\\
S_\phi & = & \frac{1}{2}\int d^4x \, \sqrt{-g} \, L(\phi,X) 
\equiv \int d^4x \, \sqrt{-g} \, G_2(\phi,X) \nonumber\\
&&\nonumber\\
&=& \frac{1}{2}\int d^4x \, \sqrt{-g} \, \left[  
A(\phi)+B(\phi)\sqrt{X(t, r)} \, \right] \,,
\end{eqnarray}
with 
\begin{eqnarray} 
A(\phi) & = & -6 \,S^2(\phi) \,,\\
&&\nonumber\\
B(\phi) & = & 4 \sqrt{2} \, S_{,\phi}(\phi) \,.
\end{eqnarray}
This action belongs to the class of $k$-essence theories of gravity and 
its  variation with respect to $\phi$ produces the well-known equation of 
motion for $\phi$ 
\begin{eqnarray}
\frac{\delta S}{\delta\phi} &=& \left( {L}_{,X} g^{\alpha\beta} - 
{L}_{,XX} \nabla^\alpha \phi \nabla^\beta \phi 
\right)\nabla_\alpha\nabla_\beta \phi- 2 X {L}_{,X \phi} \nonumber\\
&&\nonumber\\
&\, & + {L}_{,\phi}= 0 \,. \label{keom}
\end{eqnarray}
When the inequality
\begin{equation}\label{hyper}
  {L}_{,X} + 2 X {L}_{,XX} > 0 
\end{equation}
is satisfied, the partial differential equation~(\ref{keom}) is hyperbolic 
and $\phi$ describes a propagating  
degree of freedom. In the case under consideration the 
inequality~(\ref{hyper}) is violated because  
\begin{eqnarray}
L_{,X} &=& \frac{ B(\phi)}{ 2\sqrt{X} } \,,\\
&&\nonumber\\
2 X {L}_{,XX} &=& -\frac{B(\phi)}{\sqrt{X}}   \,,
\end{eqnarray}
and $ L _{,X} + 2 X {L}_{,XX}=0$, which is the condition for the scalar 
field $\phi$ to be a cuscuton 
\z\cite{Afshordi:2006ad, Afshordi:2007yx, Afshordi:2009tt, Abdalla:2013ara, 
Bhattacharyya:2016mah}. It is always 
possible to redefine 
the scalar field  $\phi\to \tilde{\phi}(\phi)$ so that 
$B(\phi) \, d\tilde{\phi}/d\phi=\pm\mu^2$ and to rewrite the cuscuton 
Lagrangian as 

\be\label{cuscuton}
L(\phi,X) = \frac{ \sqrt{-g}}{2}  
\left[ \pm\mu^2\sqrt{X}-V(\phi)\,\right]\,, 
\ee
where $V(\phi)=-A(\phi)$\,. As a consequence,  
 $ V(\phi)=6H^2(t)$ and  $ \dot{H}(t)=\mp  \, \mu^2 \,
\dot{\phi}(t)/\left( 4\sqrt{2} \right) $, 
which yield 
\begin{eqnarray} 
H(t) &=&  C \pm \frac{  \mu^2\phi(t)}{4 \sqrt{2} } \,,\\
&&\nonumber\\
V(\phi) &=& 6 \left[ C\pm  \frac{\mu^2 \phi(t)}{4\sqrt{2} } \right]^2 \,,
\end{eqnarray}
where $C$ is a constant of integration.

From the physical point of view, the cuscuton is equivalent to a perfect 
fluid and it is incompressible since its sound speed 
is found to be infinite\z\cite{Afshordi:2006ad, Afshordi:2007yx, 
Afshordi:2009tt, Abdalla:2013ara, Bhattacharyya:2016mah, 
Afshordi:2014qaa,deRham:2016ged}. This feature, which expresses the facts
that perturbations do not propagate and the cuscuton is not dynamical, is 
reminiscent of the property of the McVittie universe that the cosmic fluid 
is not allowed to accrete onto the central object but somehow expands 
rigidly, given that the energy density $\rho(t)$ is perfectly homogeneous 
\z\cite{McVittie}. This ``McVittie condition'' is removed in generalized 
McVittie geometries by allowing radial energy flux and, correspondingly, 
the class of Horndeski theories admitting generalized McVittie spacetimes
as solutions is wider than cuscuton gravity.

\section{Generalized McVittie solutions of Horndeski gravity}
\label{sec:3}
\setcounter{equation}{0}

Let us turn now to generalized McVittie geometries\z\cite{Faraoni:2007es, 
Gao:2008jv, Faraoni:2008tx} characterized by a radial spacelike energy 
flux onto (or away from) the central object, which was instead forbidden explicitly by 
McVittie in his solution of the Einstein equations\z\cite{McVittie}. The 
corresponding line element in isotropic coordinates reads
\begin{eqnarray}
ds^2 &=& - \dfrac{\left(  1-\dfrac{m(t) }{2 a(t) r} \right)^2}{\left(   
 1+\dfrac{m(t)}{2 a(t) r}  \right)^2} dt^2 \nonumber\\
&&\nonumber\\
&\, & + a^2(t) \left(1+\dfrac{m(t)}{2 a(t) r}\right)^4 \left(dr^2+r^2 
d\Omega_{(2)}^2 \right)\,, \label{genMcV}
\end{eqnarray}
where $m(t)>0 $ is now a function of time, while it was constant in the 
McVittie solution\z\cite{McVittie}. This time dependence of the mass parameter of the central object embedded in the FLRW universe is due to the non-vanishing radial energy flux\z\cite{Faraoni:2007es,Gao:2008jv, Faraoni:2008tx}, therefore generalized McVittie geometries are substantially different from the original McVittie ones. Here we determine the Horndeski theories that admit generalized McVittie solutions. The method is similar to the one adopted in the previous section: we substitute the line element \z\eqref{genMcV} in Eq.\z\eqref{hfeq} and impose that $G_2$, $G_3$, and $G_4$ solve it. Then we will compare our result with Ref.~\cite{Afshordi:2014qaa}. Let us consider again a strictly monotonic 
homogeneous scalar field $\phi(t)$ with $\dot{\phi}(t)<0$ and 
$m(t)<2a(t)r$, which leads to
\begin{equation}\label{gkin}
X(t, r) = \frac{1}{2} \, \left( \frac{ 2 \, a(t) r+m(t) }{
2 \, a(t) r-m(t) }\right)^2 \dot{\phi}^2 (t)\,,
\ee
or
\be
\dot{\phi}(t)=- \frac{ 2 \, a(t) r-m(t) }{ 2 \,a(t) r + m(t)} \sqrt{2X(t,r)} \,.
\end{equation}
The $(t,r)$ component of the Horndeski field equations yields 
\begin{equation}
\label{G3.0}
 G_{3,\phi}(\phi,X) =\frac{  G_{4,\phi}(\phi) + 2G_0(\phi) \, 
G_4(\phi) }{X} \,,
\ee
where
\be
G_0(\phi)=\dfrac{ \dot{m}(t)}{m(t) \, \dot{\phi}(t)}=\dfrac{W(t)}{\dot{\phi}(t)}\,,
\end{equation}
and with $W(t)\equiv\dot{m}(t)/m(t)$\,.\\
Equation~(\ref{G3.0}) is integrated with respect to $X$  obtaining
\begin{equation}\label{G3}
G_{3}(\phi,X)= \left[ 2 G_{0}(\phi) \,  G_4(\phi) + G_{4,\phi}(\phi)\right]\, 
{\rm ln}\, X + G_1(\phi) \,,
\end{equation}
where $G_1(\phi)$ is an arbitrary integration function of the scalar 
field which can be neglected because it produces a boundary term, as we will see. By imposing the generalized McVittie geometry~(\ref{genMcV}), the 
radial component of the field equations~(\ref{hfeq}) in conjunction with 
Eq.~(\ref{G3}) gives 
\begin{eqnarray}
G_{2}(\phi,X) &=&   \mathcal{A}(\phi,X) \, G_4(\phi) + 
\mathcal{B}(\phi,X) \, G_{4,\phi}(\phi) \nonumber\\
&&\nonumber\\
&\, & - 2 X  \left( 1- {\rm ln}\,X \right) G_{4,\phi\phi}(\phi) 
+2  X  G_{1,\phi}(\phi)\,,\nonumber\\
&& \label{G2.0}
\end{eqnarray}
where $\mathcal{A}(\phi,X)$ and $\mathcal{B}(\phi,X)$ depend on $\phi$ and 
$X$\,. Using some auxiliary functions, it is possible to write the explicit functional form of $G_2$ as

\begin{eqnarray}  
G_{2}(\phi,X) &=& \, 2 G_4(\phi) \left[ F_1(\phi)+F_2(\phi) \sqrt{X} \right.\nn\\
&&\nn\\
&\,&\left.-2  X \Big(G_{0,\phi}(\phi) \left( 2-{\rm ln}\,X \right) + 3 
G_0^2(\phi) \Big)\right] \nonumber\\
&&\nn\\
&\,& -4G_{4,\phi}(\phi) \left[ G_0(\phi) X \left(2-{\rm 
ln}\, X \right)+S_0(\phi) \sqrt{X}\right]\nn\\
&&\nn\\
&\,& - 2 X G_{4,\phi\phi} \left( 2-{\rm ln}\, X \right) + 2 X  G_{1,\phi}(\phi) \,,
 \label{G2}
\end{eqnarray}

where

\begin{eqnarray}
G_0(\phi) &=& \frac{1}{\dot{\phi}(t)} \frac{ \dot{m}(t)}{m(t)}=\frac{ W(t)}{\dot{\phi}(t)}\,,\label{g0}\\
&\,&\nonumber\\
S_0(\phi) &=& \sqrt{2} \Big( H(t)-W(t)\Big) \,, \label{s0}\\
&\,&\nonumber\\
F_1(\phi) &=& -3 \Big( H(t)-W(t)\Big)^2 =-\frac{3}{2} \, S_0^2(\phi) \,, \label{f1}\\
&\,&\nonumber\\
F_2(\phi) &=& \frac{2\sqrt{2}}{\dot{\phi}(t)}\left[\dot{H}(t)-\dot{W}(t)+3W(t)\Big( H(t)-W(t) \Big)\right]\nn\\
&\,&\nonumber\\ 
&=& 2\Big[ S_{0,\phi}(\phi)+3G_0(\phi) S_0(\phi)\Big]\label{f2}\,.
\end{eqnarray}  
These equations show that $G_1(\phi)$ is completely negligible because, together with Eq.\z\eqref{G3}, it produces the total divergence $2 X\,G_{1,\phi}(\phi)-G_1(\phi)\Box\phi = -\nabla_{\mu}[G_1(\phi)\nabla^{\mu}\phi]$ in the action integral, and therefore an irrelevant boundary term.

If $G_0(\phi)=0$, then $ m$ is constant and we recover the function $G_2$ 
associated with the  McVittie metric
\begin{eqnarray}
G_{2}(\phi,X) &=& 2 G_4(\phi) \left[ F_1(\phi)+F_2(\phi) 
\sqrt{X} \right] \nonumber\\
&&\nonumber\\
& \, & -4 G_{4,\phi}(\phi) S_0(\phi) \sqrt{X}-2 X G_{4,\phi\phi} \left( 2-{\rm ln} \,X \right) \nonumber\\
&&\nonumber\\
&\, & 
 +2 X G_{1,\phi}(\phi)\,,
\end{eqnarray}
where
\begin{eqnarray}
S_0(\phi) & = & \sqrt{2} \, H(t)\,,\\
&&\nonumber\\
F_1(\phi) &=& -3 H^2(t) = -\frac{3}{2}S_0^2(\phi) \,,\\
&&\nonumber\\
F_2(\phi) &=& 2\sqrt{2} \,\, \frac{ \dot{H}(t)}{\dot{\phi}(t)} = 2 \, S_{0,\phi}(\phi)\,.
\end{eqnarray}
In general, $G_4(\phi)$ is unconstrained.

If $m(t)$ vanishes identically, $ G_0(\phi)=0$ and one recovers the above 
set of functions with $X= \dot{\phi}(t)^2/2$\,.

The case $ \dot{\phi}(t)>0$ is obtained by changing the sign of the terms 
with $ \sqrt{X} $\,. Fixing the sign $ \dot{\phi}(t)>0 $ does not cause loss 
of generality because rewriting every quantity in terms of $t$ leads to 
the same expression for $G_2(t)$ and $G_3(t)$\,.

Therefore, the class of viable Horndeski theories that admit generalized McVittie solutions is characterized by $G_2$ and $G_3$ as in Eqs.~(\ref{G2}) and~(\ref{G3}) simultaneously with Eqs.~(\ref{g0})-(\ref{f2}), and any quadruple $ \Big\{ G_4(\phi),\phi(t),m(t),a(t)\Big\} $ corresponds to a particular solution of the Horndeski 
theory~(\ref{action}).

The results of Ref.~\cite{Afshordi:2014qaa} are recovered for 
$G_4(\phi)=1/2$; this special case in conjunction with $ 
G_3(\phi,X)=0 $ reproduces  Sec.~\ref{k-ess} with $ m =$~const.

A particular subclass of the theories found, corresponding to $G_3 = 0$, is discussed below. This subclass is of physical interest for the reasons discussed in Ref.\z\cite{Creminelli:2019kjy} and the vanishing of  $G_3$ has significant consequences.

\subsection{Vanishing \texorpdfstring{$G_3(\phi,X)$}{G3(phi,X)}}\label{G3=0}

In the special subclass of Horndeski theories with $G_3 =0$, Eq.~(\ref{G3}) gives
\begin{equation}
2 G_{0}(\phi)\, G_4(\phi)+G_{4,\phi}(\phi)=0
\ee
and
\be
G_4(\phi)= C \,\exp\left( -2\int_{1}^{\phi} G_0(\xi) \,d\xi\right) \,,
\end{equation}
where $ C$ is an integration constant. Due to the monotonicity of 
$\phi$, one can write $m(t)=\tilde{m}(\phi)$ and   $G_0(\phi)=\dfrac{d\ln\,\tilde{m}(\phi)}{d\phi}$ and, redefining the constant $C$,
\begin{equation}\label{gcoupl}
G_4(\phi)=  
\dfrac{C}{\tilde{m}^2(\phi)}
\ee
or
\be
\tilde{G}_4(t)= \dfrac{C}{{m}^2(t)}\,.
\end{equation}
Thus, $G_4 (\phi) $ is constant if and only if $m(t)$ is.

Let us restore explicitly the speed of light $c$ and the Newton coupling assuming it to be time-dependent, $G=G(t)$\,. In this way, the McVittie mass coefficient can be rewritten as
\begin{equation}\label{littlem}
m(t) = \dfrac{G(t)M(t)}{c^2}\,,
\end{equation}
where the function $M(t) $ has the dimensions of a mass, while the Horndeski nonminimal coupling reads
\begin{equation}\label{g4coupling}
    G_4(\phi)=\dfrac{c^4}{16\pi G(t)}\,.
\end{equation} 
Thus, Eq.~(\ref{gcoupl}), together with Eqs.\z\eqref{littlem} and\z\eqref{g4coupling}, can be written as 
\begin{eqnarray}
G(t)&=& G_N \, \dfrac{M_0^2}{M(t)^2}\,,\label{G&M1}
\end{eqnarray}
and it results
\begin{eqnarray}
m(t) &=& \dfrac{G_N 
M_0}{c^2} \, \dfrac{M_0}{M(t)} \label{G&M2} 
\end{eqnarray}
that, in turn, give
\be
\frac{ \dot{m}(t)}{m(t)}=-\dfrac{ \dot{M}(t)}{M(t)}\,,\label{G&M3}
\end{equation}
where $G_N$ is the present value of the  gravitational coupling and $M_0\equiv M(t_0) $  is the present value of $M(t)$\,.

\section{Horndeski--McVittie as an extended cuscuton model}
\label{sec:4}
\setcounter{equation}{0}

Here we show that the class of viable Horndeski theories admitting McVittie Eq.\z\eqref{McVittie} and generalized McVittie Eq.\z\eqref{genMcV} solutions is a particular case of what in the literature is called extended cuscuton model\z\cite{Iyonaga:2018vnu,Iyonaga:2020bmm}, a generalization of the cuscuton seen in Sec.~\ref{k-ess} in which the scalar $\phi$ remains non-dynamical.

The cuscuton\z\eqref{cuscuton} has always received much attention (\cite{Afshordi:2006ad, Afshordi:2007yx, Afshordi:2009tt, Abdalla:2013ara, Bhattacharyya:2016mah, 
Afshordi:2014qaa,Iyonaga:2018vnu}, see also Refs.\z\cite{deRham:2016ged, 
Boruah:2017tvg, Boruah:2018pvq, Romano:2016jlz, Chagoya:2016inc, 
Andrade:2018afh, Ito:2019fie, Ito:2019ztb}), representing (in the 
unitary gauge, $\phi=\phi(t)$) the unique 
subclass of $k$-essence theory with only two propagating degrees of 
freedom. This fact is related to two closely related features of this model:

\begin{itemize}

\item  The equation of motion for the scalar $\phi$ is of first 
order in the case of FLRW cosmology; 

\item The kinetic term of scalar cosmological perturbations vanishes.

\end{itemize}
 
However, the cuscuton is not the most general Horndeski theory propagating only two degrees of freedom. The theory with this feature is the extended cuscuton model, characterized by the generalization of the first statement to 

\begin{itemize}

\item The system composed of the dynamical equations\z(\ref{hfeq}) and\z(\ref{heom}) is degenerate in FLRW (see Ref.\z\cite{Iyonaga:2018vnu}).

\end{itemize}

The above statement implies that the 
functions $G_i(\phi, X)$ in the action~(\ref{action}) must satisfy the 
conditions
\begin{eqnarray}    
G_2(\phi,X) &=& f_1(\phi) + f_2(\phi) \sqrt{2X} \nonumber\\
&&\nonumber\\
&\, & -\left( 2f_{3,\phi}(\phi)+4f_{4,\phi\phi}(\phi)+\frac{3{f_3}^2(\phi)}{4f_4(\phi)}\right)X \nn\\
&&\nonumber\\
&\, &  + \Big( f_{3,\phi}(\phi)+2f_{4,\phi\phi}(\phi)\Big)  
X \ln X \,, \label{c2}\\
&&\nonumber\\
 G_3(\phi,X) &=& \left( \frac{1}{2}f_3(\phi)+f_{4,\phi}(\phi)\right) 
\ln{X }  \,, \label{c3}\\
&&\nonumber\\
G_4(\phi) &=& f_4(\phi) \,, \label{c4}
\end{eqnarray}
where the $ \{f_i\}$ are arbitrary functions of $\phi(t)$, so that the 
scalar field does not propagate at  both the background and the  
perturbative levels.

Eqs.~(\ref{c2})-(\ref{c4}) coincide with Eqs.~(\ref{G3})-(\ref{G2}) under the  identifications
\begin{eqnarray}
   f_1(\phi) &=& 2 \, G_4(\phi)F_1(\phi)=-3 G_4(\phi) S_0^2(\phi)\,,\label{cf1}\\
&&\nonumber\\
f_{2}(\phi) &=& \sqrt{2} \, \Big[ G_4(\phi)F_2(\phi)+2S_0(\phi)\,  G_{4,\phi}(\phi)\Big]\nonumber\\
            &&\nonumber\\
            &=& 2\sqrt{2} \, \Big[G_4(\phi)(S_{0,\phi}(\phi)+3G_0(\phi) S_0(\phi))\nn\\
            &&\nonumber\\
            &\, & +S_0(\phi) G_{4,\phi}(\phi)\Big] \,,\label{cf2}\\
&&\nonumber\\ 
f_{3}(\phi) &=& 4 \, G_0(\phi) \, G_4(\phi)\,,\label{cf3}
\end{eqnarray}
where 
\begin{eqnarray}
    G_0(\phi) &=& \,\dfrac{W(t)}{\dot{\phi}(t)}\,,\\
    &&\nonumber\\
    S_0(\phi) &=& \sqrt{2} \, \Big(H(t)-W(t)\Big)\,.
\end{eqnarray}

Fixing $f_3(\phi)=0$, or $ G_0=0$, Eqs.~(\ref{c2}),\z(\ref{c3}), 
(\ref{c4}),\z(\ref{cf1})-(\ref{cf3}) are equivalent to 
the set of functions\z\eqref{g2} and~\eqref{g3}. Therefore, also 
the standard ({\em i.e.}, $m=$~const.) McVittie geometry in Horndeski 
theory represents an extended cuscuton model. 
It is well-known that this kind of theory is transformed into the 
standard cuscuton model\z\cite{Afshordi:2006ad, Afshordi:2007yx, 
Afshordi:2009tt, Abdalla:2013ara, Bhattacharyya:2016mah}
by a particular disformal transformation\z\cite{Afshordi:2014qaa,Iyonaga:2018vnu}.

Extended cuscuton models are used to describe dark energy, mimicking the $\Lambda$-CDM model in late-time cosmology\z\cite{Iyonaga:2020bmm,Maeda:2022ozc} with a time dependent gravitational coupling, or as a mechanism to remove the cosmological singularity through a bounce\z\cite{Quintin:2019orx}. These works deal with a particular choice of the functions $\{f_i\}$ which provide different cosmological models.
In Refs.\z\cite{Iyonaga:2020bmm,Maeda:2022ozc,Quintin:2019orx}, they discuss the extended cuscuton action in the presence of matter while, until now in the present work we have considered only the vacuum Horndeski action ({\em i.e.}, $G_2$ and $G_3$ describe {\em effective} matter). Therefore, we are able to find a functional form of the viable Horndeski action such that the McVittie geometry satisfies the field equations but there is freedom in choosing the functions\z\eqref{g0} and\z\eqref{s0} and one cannot discuss the dynamics of the system until these are fixed. 
For this reason, in the next section we improve the generality by adding a matter fluid to the extended cuscuton model,Eqs.\z\eqref{c2}-\eqref{c4} and classifying all the resulting possibilities.

\section{McVittie in the extended cuscuton model with matter}
\label{sec:5}
\setcounter{equation}{0}

Without loss of generality, let us parametrize the functions $\{f_i\}$ of Eqs.\z\eqref{c2}-\eqref{c4} in the more convenient  way
\begin{align}
    f_1(\phi)&\to2f_1(\phi)G_4(\phi)\,,\\
    \nonumber\\
    f_2(\phi)&\to\sqrt{2}f_2(\phi)G_4(\phi)\,,\\
    \nonumber\\
    f_3(\phi)&\to4f_3(\phi)G_4(\phi)\,.
\end{align}
This is just a redefinition of the functions $\{f_i\}$ because $f_1$, $f_2$, and $G_0$ are still free.
Moreover, we assume the cosmic fluid to be described by an imperfect fluid stress-energy tensor $T_{\mu\nu}$, allowing {\em a priori} a non-vanishing radial flow,
\begin{equation}
    T_{\mu\nu}=(P+\rho)u_\mu u_\nu+Pg_{\mu\nu}+q_\mu u_\nu+ q_\nu u_\mu \,,
\end{equation}
where the fluid four-velocity $u^\mu$ is normalized to $u^\mu u_\mu=-1$, the purely spatial vector $q^\mu$ describes a radial energy flow, and $\rho(t,r)$ and $P(t,r)$ denote the energy density and pressure of the fluid, respectively. It follows that 
\begin{equation}
    T^{t}{}_{t}=-\rho(t,r)\,,\quad T^{r}{}_{r}=P(t,r)\,.
\end{equation}
Now, the $(t,r)$ field equation gives
\begin{equation}
    f_3(\phi)=\dfrac{1}{\dot{\phi}(t)}\left[W(t)+\dfrac{\Big(2  a(t)r-m(t)\Big)^3\,T^{t}{}_{r}(t,r)}{8 a(t) m(t) G_4(\phi) \Big( 2 a(t)r + m(t)\Big)} \right]\,,
\end{equation}
where $W(t)=\dot{m}(t)/m(t)$\,. Since, by definition, $f_3$ is a function of $\phi(t)$, we can introduce a generic function of time $Q(t)$ and rewrite $T^{t}{}_{r}$ as
\begin{equation}
    T^{t}{}_{r}=\dfrac{8 a(t) \, m(t) \, G_4(\phi) \Big( 2a(t)r + m(t) \Big)}{\Big( 2  a(t)r - m(t)\Big)^3 }\, Q(t)\,,
\end{equation}
The flux $T^{t}{}_{r}$ vanishes if $m(t)=0$, and also as $r\to +\infty$\,. 
Therefore, we have
\begin{equation}
    f_3(\phi)=\frac{W(t)+Q(t)}{\dot{\phi}(t)}\,.
\end{equation}
Using this form of $f_3(\phi)$, we must now impose that the equation of motion Eq.\z\eqref{heom} of $\phi$ is satisfied, thus obtaining constraints on $f_1$ and $f_2$ that can be used in the field equations. However, it is not possible to follow this procedure for arbitrary $G_4(\phi)$, $Q(t)$, $W(t)$, and $H(t)$, and it is necessary to make some assumptions on these four functions in order to solve the algebraic expression coming from the equation of motion of $\phi$\,. This situation is related with the fact that this equation of motion sets constraints on $f_1$ and $f_2$ which are obtained in specific domains. We provide a classification of all the possible cases in which Eq.\z\eqref{heom} is satisfied by reasoning on $\Big\{G_{4,\phi}(\phi),Q(t),W(t),H(t)\Big\}$ according to whether one or more of them vanishes identically. To anticipate the results, when  $Q(t)\neq0$, $W(t)=0$, and $H(t)\neq0$, the equation of motion cannot be satisfied, as well as as when $G_{4,\phi}(\phi)\neq0$, $Q(t)\neq0$, $W(t)\neq0$, and $H(t)=0$\,. All possible cases are summarized in the following list:
\begin{itemize}
    \item Case $1$: $Q(t)=0$ and $W(t)=0$\,,
    \item Case $2$: $Q(t)=0$ and $W(t)\neq0$\,,
    \item Case $3$: $Q(t)\neq0$, $W(t)=0$, and $H(t)=0$\,,
    \item Case $4$: $G_{4,\phi}=0$, and $Q(t)=-W(t)\neq0$\,,
    \item Case $5$: $G_{4,\phi}=0$, $Q(t)\neq0$ and $W(t)=H(t)\neq0$\,,
    \item Case $6$: $G_{4,\phi}\neq0$, $Q(t)\neq0$ and $Q(t)\neq-H(t)$, and $W(t)=H(t)\neq0$\,,
\end{itemize}
where each case is characterized by the constraints on $\Big\{G_{4,\phi}(\phi),Q(t),W(t),H(t)\Big\}$\,.

\subsection{Case 1: \texorpdfstring{$Q(t)=0$}{Q(t)=0} and \texorpdfstring{$W(t)=0$}{W(t)=0}} 

We consider a generic function $G_4( \phi)$, which could be constant, vanishing radial flow, and a standard McVittie geometry. This situation occurs also in the limit of asymptotically flat spacetime $H(t)=0$\,. From the equation of motion for $\phi$ one obtains
\begin{align}
    &f_{1,\phi}(\phi)= -\big[f_1(\phi)-3 H^2(t)\big]\frac{ G_{4,\phi}(\phi)}{G_{4}(\phi)}-\frac{3}{\sqrt{2}}H(t)f_2(\phi)\,,\\
    &f_2=f_2(\phi)\,,
\end{align}
leaving the function $f_2$ unconstrained. The corresponding field equations provide the energy density and the pressure of the fluid
\begin{align}
    \rho(t)=& \, G_4(\phi) \left[f_1(\phi) + 3 H^2(t)\right]\,,\\
    \nonumber\\
    P(r,t)=&-G_4(\phi) \bigg[f_1(\phi) + 3 H^2(t)\nn\\
    \nn\\
    &+  \frac{2a(t)r + m}{ 2a(t)r- m }  \left(\sqrt{2} \, f_2(\phi)\, \dot{\phi}(t)-4\,  \dot{H}(t)\right)\bigg]\nn\\
    \nn\\
    &- 2\, \frac{ 2 a(t)r +m}{ 2a(t)r - m} \,   H(t) \, \dot{\phi}(t) \, G_{4,\phi}(\phi) \,.
\end{align}

\subsection{Case 2: \texorpdfstring{$Q(t)=0$}{Q(t)=0} and \texorpdfstring{$W(t)\neq0$}{W(t)!=0}} 
In this case there is no radial flow and there is a central inhomogeneity with non-constant mass parameter: the scalar field \av compensates\cv the time variation of $m(t)$ as it happens in the case of generalized McVittie geometries in the vacuum extended cuscuton model.

The equation of motion of the scalar field yields the constraints
\begin{eqnarray}
    f_{1,\phi}(\phi)&=&-\left[f_1(\phi)+3\Big(H(t)-W(t)\Big)^2\right]\dfrac{G_{4,\phi}(\phi)}{G_{4}(\phi)}\nn\\
    &&\nn\\
    &\,&-\frac{6}{\dot{\phi}(t)}\Big( H(t)-W(t) \Big)\left(\dot{H}(t)-\dot{W}(t) \right) ,\\
    \nn\\
    f_2(\phi)&=& \, \dfrac{2\sqrt{2}}{\dot{\phi}(t)} \, \bigg[\dot{H}(t)-\dot{W}(t)+3W(t)\Big(H(t)-W(t)\Big)\nn\\
    &\,&+\Big(H(t)-W(t)\Big)\dfrac{\dot{\phi}(t)G_{4,\phi}(\phi)}{G_{4}(\phi)}\bigg]\,.
\end{eqnarray}
As a consequence, from the field equations we obtain that the only possible cosmological fluid is a perfect fluid with equation of state parameter $-1$:
\begin{align}
    P(t)=&-G_{4}(\phi)\left[f_1(\phi)+3\Big(H(t)-W(t)\Big)^2\right]\nn\\
    \nn\\
    =&-\rho(t)\,.
\end{align}

In this case $f_2$ has a fixed functional form, which corresponds to Eq.\z\eqref{cf2} and remains constrained in the limit $W(t)\to0$\,. In general, it does not give back the previous case. 

This result is consistent with what we obtained in the vacuum case. Indeed, for vanishing $P$ and $\rho$, it is $f_1=3[H(t)-W(t)]^2$, corresponding to Eq.\z\eqref{cf1}.

\subsection{Case 3: \texorpdfstring{$Q(t)\neq0$}{Q(t)!=0}, \texorpdfstring{$W(t)=0$}{W(t)=0}, and \texorpdfstring{$H(t)=0$}{H(t)=0}}

This case describes a Schwarzschild black hole embedded in an imperfect fluid with radial flow.
The scalar field equation of motion gives
\begin{align}
    f_1(\phi)=&\frac{\lambda}{G_4(\phi)}\,,\quad f_2=f_2(\phi)\,,\\
    \nn\\
    Q(t)=&\frac{Q_0}{\sqrt{G_4(\phi)}}\,,
\end{align}
where $\lambda$ and $Q_0$ are integration constants. The fluid is characterized by
\begin{align}
    \rho(t,r)=& \, \lambda+3\, \left( \frac{2a(t)r + m}{2a(t)r -m} \right)^2 \, Q_0^2\,,\\
    \nn\\
    P(t, r)=&-\lambda+3 \,\left( \frac{2a(t)r +m}{2a(t)r-m}\right)^2 \, Q_0^2\nn\\
    \nn\\
    &+\frac{1}{\sqrt{2}} \, \frac{2a(t)r+m}{2a(t)r-m} \, f_2(\phi) \, \dot{\phi}(t) \, G_4(\phi)\nn\\
    \nn\\
    &+ \left( \frac{2a(t)r+m}{2a(t)r-m} \right)^2 \, \frac{G_{4,\phi}(\phi) \, \dot{\phi}(t)}{\sqrt{G_4(\phi)}} \, Q_0\,.
\end{align}

\subsection{Case 4: \texorpdfstring{$G_{4,\phi}=0$}{G4'(phi)=0}, and \texorpdfstring{$Q(t)=-W(t)\neq0$}{Q(t)=-W(t)!=0}}

Now we have a generalized McVittie metric with generic scale factor and constant gravitational coupling. The solution of the equation of motion of $\phi$ is
\begin{align}
    f_{1,\phi}(\phi)=0\,,\quad f_2(\phi)=0\,,\quad Q(t)=-W(t)\,,
\end{align}
where the last equality $Q(t)=-W(t)$ implies $f_3=0$ and, in conjunction with $G_{4,\phi}=0$, it has the consequence that $G_3=0$\,.

The energy density and pressure of the imperfect fluid surrounding the central object are
\begin{align}
   \rho(t,r) =& \, G_4\left[f_1+3\left(H(t)+\frac{2m(t)}{2a(t)r-m(t)}  W(t)\right)^2\right]\,,\\&\nn\\
   P(t, r)=& \, G_4\bigg[-f_1-3H^2(t)\nn\\
   &\nn\\
   &+\frac{4m(t)\Big( 3m^2(t)-8m(t)a(t)r -4a^2(t) r^2\Big)}{ \Big( 2a(t)r-m(t)\Big)^3} \, W^2(t)\nn\\
   &\nn\\
   &-\frac{4m(t)\Big( 4a(t)r-m\Big) \Big( 2a(t)r-3m(t) \Big)}{\Big( 2a(t)r-m(t)\Big)^3} \, W(t)H(t)\nn\\
   &\nn\\
   &-\frac{4m(t)\Big( 2a(t)r + m(t) \Big)}{\Big( 2a(t)r-m(t)\Big)^2} \, W'(t)\nn\\
   &\nn\\
   &-2 \, \frac{2a(t)r-m(t)}{2a(t)r-m(t)} \, H'(t)\bigg]\,.
\end{align}
The homogeneous limit $r\to\infty$ reproduces the usual perfect fluid Friedmann equation
\begin{align}
    \rho(t)\simeq& \, G_4 \Big[ f_1+3H^2(t) \Big]\,,\\
    \nn\\
    P(t)\simeq&-G_4 \left[ f_1+3H^2(t)-2\dot{H}(t) \right] \,,
\end{align}
where $f_1$ plays the role of a cosmological constant. 
Therefore, this case is equivalent to the  generalized McVittie metric of  GR with a cosmological constant.

\subsection{Case 5: \texorpdfstring{$G_{4,\phi}=0$}{G4'(phi)=0}, \texorpdfstring{$Q(t)\neq0$}{Q(t)!=0} and \texorpdfstring{$W(t)=H(t)\neq0$}{W(t)=H(t)}}

The condition $W(t)=H(t)$ automatically makes the generalized McVittie solution a non-rotating Thakurta geometry which we are going to discuss  below. 
When one  solves the equation of motion of $\phi$ imposing $G_{4,\phi}=0$, $Q(t)\neq0$, $W(t)\neq0$, and $H(t)\neq0$ and assuming $Q(t)\neq-W(t)$, one obtains
\begin{align}
    W(t)&=H(t)\quad\Rightarrow\quad m(t)=m_0 \, a(t)\,,\\
    \nn\\
    Q(t)&=\frac{Q_0}{a^3(t)}\,,
\end{align}
where $m_0$ and $Q_0$ are integration constants. In an expanding universe, the equation $W(t)=H(t)$ implies a growing black hole mass. In addition, the condition $Q(t)\neq-W(t)$ becomes $Q(t)\neq-H(t)$\,.

In this case, the cosmic fluid has energy density and pressure
\begin{align}
    \rho(t,r) = & \, G_4\left[3\left(\dfrac{2a(t)r+m(t)}{2a(t)r-m(t)}\right)^2 \, \dfrac{Q_0^2}{a^6(t)}+f_1\right]\,,\\
    \nn\\
    P(t,r)= & \, G_4\left[3\left(\dfrac{2a(t)r+m(t)}{2a(t)r-m(t)}\right)^2  \dfrac{Q_0^2}{a^6(t)}-f_1\right]\,.
\end{align}
As $r \to +\infty$, the cosmic fluid becomes a perfect fluid and $\rho$ and $P$ become
\begin{align}
    \rho(t) \simeq & \, G_4\left( \frac{3 \, Q_0^2}{a^6(t)}+f_1\right)\,,\\
    \nn\\
    P(t) \simeq &  \, G_4\left( \frac{ 3 \, Q_0^2}{a^6(t)}-f_1\right)\,,
\end{align}
with $f_1$ acting again as a cosmological constant. Therefore, the cosmic fluid is made up by a cosmological constant contribution and by a stiff fluid with linear barotropic coefficient equal to 1.

\subsection{Case 6: \texorpdfstring{$G_{4,\phi}\neq0$}{G4'(phi)!=0}, \texorpdfstring{$Q(t)\neq0$}{Q(t)!=0} and \texorpdfstring{$Q(t)\neq-H(t)$}{Q(t)!=-H(t)}, and \texorpdfstring{$W(t)=H(t)\neq0$}{W(t)=H(t)!=0}}

If none of $G_{4,\phi}(\phi)$, $Q(t)$, $W(t)$, $H(t)$ vanishes, the equation of motion of the scalar field yields
\begin{align}
    f_1(\phi)=&\frac{\lambda}{G_4(\phi)}\,,\quad f_2(\phi)=0\,,\\
    \nn\\
    W(t)= & H(t)\,,\quad\quad Q(t)\neq-H(t)\,,\\
    &\nn\\
    \dot{Q}(t)=&-3 \, Q(t) \, H(t)\nn\\
    \nn\\
         &-\frac{Q(t)}{2}\,\frac{\dot{\phi}(t) \, G_{4,\phi}(\phi) }{G_4(\phi) }\left(\dfrac{ Q(t)+2H(t) }{Q(t)+H(t)}\right)\nn\\
         \nn\\
    =&-3 \, Q(t)H(t)-\frac{Q(t)}{2}\,\frac{\dot{G}_{4}(t) }{G_4(t) }\left(\dfrac{ Q(t)+2H(t) }{Q(t)+H(t)}\right)\,,\quad\label{Qdot}
\end{align}
where $\lambda$ is an integration constant. 

 In this case $W(t)=H(t)\Rightarrow m(t)=m_0 \, a(t)$ does not mean that an expanding universe corresponds to a growing mass, because $m(t)$ is a mass parameter  containing a non-constant gravitational coupling, then we have to take into account the whole evolution of the function which is not a mere change of mass.
 
 Indeed, with the parameterization of Sec.\z\ref{G3=0}, it is
\begin{equation}
    \dfrac{\dot{M}(t)}{M(t)}-\dfrac{\dot{G_4}(t)}{G_4(t)}=H(t)\,,
\end{equation}
 while Eq.\z\eqref{Qdot} can be seen as a sort of continuity equation. Then, the field equations give
\begin{align}
    \rho(t,r)=& \, \lambda +\,3G_4(\phi) \, Q^2(t) \,\left(\frac{2a(t)r + m(t) }{ 2a(t)r+m(t)}\right)^2,\\
    \nn\\
    P(t,r)=&-\lambda+\bigg[ 3\, G_4(t) \, Q^2(t)+\frac{ \, \dot{G}_{4}(t) \, Q^2(t)}{Q(t)+H(t)}\bigg]\times\nn\\&\nn\\
    &\times\left(\frac{2a(t)r + m(t) }{ 2a(t)r+m(t)}\right)^2 \,.
\end{align}
The homogeneous limit $r\to\infty$ reduce these quantities to
\begin{align}
    \rho(t)& \simeq \lambda+3G_4(t)Q^2(t)\,,\\
    \nn\\
    P(t)& \simeq -\lambda+3G_4(t) \, Q^2(t)+\frac{\dot{G}_{4}(t)\, Q^2(t)}{Q(t)+H(t)}\,.
\end{align}
Even if, to solve the scalar field equation, we impose the non-vanishing of $G_{4,\phi}(\phi)$, at the end of the day this case generalizes the previous one. Indeed, imposing $G_{4,\phi}(\phi)=0$ one obtains the same Friedmann equations and the same expression for $Q(t)$ in terms of $a(t)$\,.

The current case is the most interesting one: to investigate it, let us consider a matter-dominated cosmological era and let us assume the barotropic equation of state $P(t)=w \, \rho(t)$,  $w=$~const. In particular, for dust, we can neglect $\lambda$ and  obtain
\begin{align}
    H(t)&=-Q(t)-\frac{\dot{G}_4(t)}{3\, G_4(t)}\,,\\
    \nn\\
    \dot{Q}(t) & = \frac{3}{2} \, Q^2(t)\quad\Rightarrow\quad Q(t)= -\frac{2}{2 c_1+3 t}\,,
\end{align}
where $c_1$ is an integration  constant, and 
\begin{align}
    G_4(t)&= \, \frac{ c_2 \left( 2 c_1+3 t \right)^2}{a^3(t)}\,,\\
    \nn\\
    \rho(t) & = \frac{12 \, c_2}{a^3(t)}\,,
\end{align}
\begin{align}
     M(t)&=\, \frac{ c_3\left( 2 c_1+3 t \right)^2}{a^2(t)}\,,
\end{align}
where $c_{2,3}$ are integration constants. Therefore, the density of dust scales as  $\rho\propto a^{-3} $ as in GR. We can generalize this treatment to a generic equation of state constant parameter $w$\,. Assuming the linear barotropic equation of state $P(t)=w\rho(t)$ leads to the system of equations (when $w\neq 1$ and $\dot{G}_4(t)\neq0$)
\begin{align}
   \frac{\dot{G}_4(t)}{G_4(t)}=& \, 3 \left( w-1 \right) \Big( H(t)+Q(t) \Big)\,,\\
   \nn\\
    \dot{Q}(t)=&-\frac{3 Q(t)}{2} \Big( 2 w H(t) + (w-1) Q(t) \Big) \,,
\end{align}
yielding
\begin{align}
    Q(t)= &  \dfrac{a^{-3 w}(t)}{c_1+\dfrac{3}{2}\displaystyle\int_1^t  (w-1) a^{-3 w}(\tilde{t}) \, d\tilde{t}}\,,\\
    \nn\\
    G_4(t)=  \, & c_2 \, a^{3(w-1)} \left[ c_1+\dfrac{3}{2}\int_1^t (w-1) \, a^{-3 w}(\tilde{t}) \, d\tilde{t}\right]^2,
\end{align}
with $c_{1,2}$ are integration constants. Then, the energy density is
\begin{align}
    \rho(t)=&\dfrac{3c_2}{ a^{3(1 + w)} (t)}\,,
\end{align}
while the  black hole mass is
\begin{align}
    M(t)=&c_3 \, a^{3 w-2}(t)\left[ c_1+ \dfrac{3}{2}\int_1^t (w-1) a^{-3 w}(\tilde{t}) \, d\tilde{t}\right]^2 \,,
\end{align}
where $c_3$ is another integration constant and the scale factor is unknown. At first sight  this result seems to hold also for $w=-1$, the simplest dark energy model, and for $w=1$, the stiff matter case, but the latter value of the equation of state parameter is forbidden and can be taken into account only in the case $\dot{G}_{4}=0$ (the previous case).

It is worth asking   what happens if the change in the effective mass is due exclusively to the change of gravitational coupling. Adopting  the parametrization of Sec.\z\ref{G3=0}, $W(t)=-\dot{G}_4(t)/G_4(t)\rightarrow \dot{G}_4(t)/G_4(t)=-H(t)\Rightarrow G_4(t)=c_G/a(t)$\,. The consequence is that
\begin{align}
    Q(t)=& -\frac{(3 w-2)}{3 (w-1) } \, H(t)\,,\\
    \nn\\
    H(t)=&\frac{2}{t (3 w+2)-2 c_1}\,,\\
    \nn\\
    a(t)=& c_2 \left[ t (3 w+2)-2 c_1\right]^{\frac{2}{3 w+2}}\,,\\
    \nn\\
    \rho(t)=&\frac{4 c_G (2-3 w)^2}{3 c_2 (w-1)^2}\left[ t (3 w+2)-2 c_1\right]^{-\frac{6 (w+1)}{3 w+2}}\,,
\end{align}
where $\{c_i\}$ are  integration constants. Therefore, in this model, the barotropic coefficient is constrained to values $w>-2/3$\,.

A final comment about the condition $W(t)=H(t)\Rightarrow m(t)=m_0 \, a(t)$ is mandatory: when the latter holds, the generalized McVittie line element assumes the  form
\begin{align}
    ds^2 =& {-\dfrac{\left(1-\dfrac{m_0}{2 r }\right)^2}{\left(1+\dfrac{m_0}{2 r}\right)^2}} \, dt^2 \nn\\&+  a^2(t)\left(1+\dfrac{m_0}{2 r}\right)^4\left(dr^2+r^2 d\Omega_{(2)}^2\right)\,.\label{newM}
\end{align}
 This special case of  generalized McVittie metrics is the non-rotating Thakurta solution\z\cite{Thakurta}, see also Refs.\z\cite{Mello:2016irl, Carrera:2009ve, Faraoni:2009uy}, which is conformal to the Schwarzschild metric  
with conformal factor $a(\eta)$, where $\eta$ is the conformal time of the FLRW ``background'' defined by $dt=ad\eta$\,. Using the \av conformal Schwarzschild radius\cv
\begin{equation}
    R=r\left(1+\dfrac{m_0}{2r}\right)^2\,,
\end{equation}
 as the radial coordinate, the metric\z\eqref{newM} can be rewritten as 
\begin{align}\label{Tmetric}
    ds^2 =&-\left(1-\frac{2m_0}{R}\right)dt^2+a^2(t)\left(1-\frac{2m_0}{R}\right)^{-1}dR^2\nn\\
    &\nn\\
    &+a^2(t)R^2d\Omega_{(2)}^2 \nn\\
    &\nn\\
    = &a^2(\eta)\Bigg[-\left(1-\frac{2m_0}{R}\right)d\eta^2+\left(1-\frac{2m_0}{R}\right)^{-1}dR^2\nn\\
    &\nn\\
    &+R^2d\Omega_{(2)}^2\Bigg]\,.
\end{align}
The metric\z\eqref{newM} has a pleasant characteristic which is missing in the standard McVittie metric: it provides the usual Hubble law $v=H(t)d$\,. Unlike the McVittie universe, the Hubble law holds exactly and not just asymptotically. This aspect prevents the possibility that the current $H_0$ tension be due to proximity to an inhomogeneity, at least in this model. This could be a straightforward theoretical solution for the Hubble tension problem.

\section{Discussion and Conclusions}
\label{sec:10}
\setcounter{equation}{0}

There are several  motivations to revisit McVittie and generalized McVittie solutions in the context of  Horndeski gravity, after the progresses recently made  in the study of these theories. Furthermore, there are now theoretical indications for, and the experimental capability  to detect, scalar hair around black holes. This hair could vary on cosmological scales, and the McVittie solutions of scalar-tensor gravity offer a context for this possibility. While it was shown that McVittie geometries solve cuscuton theory and generalized McVittie geometries solve more general Horndeski theories, previous studies were limited to vacuum. We have now included matter in the picture, which is essential in order to achieve a realistic cosmological background where to embed the McVittie central object. Moreover, the extended cuscuton scenario introduced very recently has not been considered before in relation with (generalized) McVittie spacetimes. As in the ``old'' cuscuton, the scalar field of the extended cuscuton does not propagate  degrees of freedom. Here we  found conditions under which generalized McVittie geometries are solutions of the extended cuscuton model.

The requirement that the scalar field be homogeneous is essential for the
McVittie and generalized McVittie geometries to be solutions of  
the Horndeski field equations. Without this property, the field equations, 
rewritten as effective Einstein equations, exhibit an effective imperfect 
fluid stress-energy tensor in their right-hand sides\z\cite{Pimentel89, 
Faraoni:2018qdr}. The  characteristic quantities of an imperfect fluid, 
{\em i.e.}, heat current density, shear, and viscous pressure are due to  the nonminimal coupling of $\phi$ with gravity\z\cite{Pimentel89, Faraoni:2018qdr} and preclude (generalized) McVittie from solving the field equations because these geometries are shear-free 
\z\cite{KustaanheimoQvist, Raychaudhuri}. The (generalized) McVittie 
geometry is inhomogeneous, due to the presence of the central object. This  inhomogeneity dies off asymptotically as the metric becomes FLRW at large  distances from the central object. Therefore, the homogeneous scalar field 
does not affect directly the gravitational field of the central object  because it sources only the FLRW ``background'' universe in which the 
latter is embedded and, in the generalized McVittie case in vacuo, it determines the energy current onto it. In both cases, $\phi(t)$ controls the evolution of the mass parameter $m(t)$\,. It can legitimately be said that the scalar $\phi(t)$ is ``locally stealth'' on small scales near the 
central object but is not stealth with respect to the large-scale FLRW 
universe (indeed, it sources it). In this context, we  restricted ourselves to the subclass of 
viable Horndeski theories. 

 We  found that, in the extended cuscuton model in  presence of matter, a special generalized McVittie geometry appears as a solution, the non-rotating Thakurta metric. This geometry is peculiar since, contrary to all other (generalized) McVittie solutions,  it is conformally equivalent to the Schwarzschild one where  $m(t)=m_0 a(t)$ is scaling as a length in the FLRW background.  More important, it was shown in Ref.\z\cite{Gao:2008jv} that the generalized McVittie solutions of general relativity with a fluid have the non-rotating Thakurta geometry as a late-time attractor. The proof of this statement in Ref.~\cite{Gao:2008jv} depends only on the functional form of the metric and not on the field equations, and can be transposed without change to the extended cuscuton model, provided that a fluid is present. Therefore, non-rotating Thakurta becomes the generic solution at late times in the class of generalized McVittie geometries. We stress that the presence of a fluid is essential: in Horndeski models without matter, where the scalar field acts  as the fluid for the FLRW background,  the late-time limit may make the scalar $\phi$ constant and make the fluid disappear. In this case, one should not expect a non-rotating Thakurta limit for the generalized McVittie solution. Indeed, this is the case for the generalized McVittie solution of the ``cuscuta-Galileon'' theory found in Ref.~\cite{Afshordi:2014qaa} which could have other interesting aspects, also from an observational point of view, in the  Galileon model\z\cite{Salzano}. A more general study of the late-time behaviour of the solutions for various specific models is deferred to future work. 

Several aspects need to be analyzed in greater detail, such as the presence of late-time attractors for (generalized) McVittie spaces, other geometries describing objects embedded in cosmological spacetimes, anisotropic fluids, and the use of McVittie toy models to explore the current Hubble tension problem in relation with local inhomogeneities. These subjects will be examined in 
future research.

\begin{acknowledgments} 
M.M., D.V., and S.C. acknowledge the support of Istituto Nazionale di Fisica Nucleare (INFN) {\it iniziative specifiche} TEONGRAV, QGSKY, and MOONLIGHT2. V.F. is supported by the Natural Sciences \& Engineering 
Research Council of Canada (grant no. 2016-03803).

\end{acknowledgments}




\end{document}